\documentclass[twocolumn,prb,amssymb,amsmath,superscriptaddress,showpac]{revtex4}
\usepackage{graphicx}
\input{ulem.sty}

\begin{document}

\title{Renormalization of hole-hole interaction at decreasing Drude
conductivity}

\author{G.~M.~Minkov}
\author{A.~A.~Sherstobitov}
\affiliation{Institute of Metal Physics RAS, 620219 Ekaterinburg,
Russia}
\author{A.~V.~Germanenko}
\author{O.~E.~Rut}
\affiliation{Institute of Physics and Applied Mathematics, Ural
State University, 620083 Ekaterinburg, Russia}

\author{B.~N.~Zvonkov}
\address{Physical-Technical Research Institute, University of
Nizhni Novgorod, 603600 Nizhni Novgorod, Russia}

\date{\today}

\begin{abstract}
The diffusion contribution of the hole-hole interaction to the
conductivity  is analyzed in gated GaAs/In$_x$Ga$_{1-x}$As/GaAs
heterostructures. We show that the change of the interaction
correction to the conductivity with the decreasing Drude
conductivity results both from  the compensation of the singlet and
triplet channels and from the arising prefactor $\alpha_i<1$ in the
conventional expression for the interaction correction.
\end{abstract}

\maketitle

The quantum corrections to the conductivity in disordered metals and
doped  semiconductors are intensively studied since
1980.\cite{Altshuler} Two mechanisms lead to these corrections: (i)
the interference of the electron waves propagating in opposite
directions along closed paths (WL correction); (ii)
electron-electron ({\it e-e}) or hole-hole ({\it h-h}) interaction.

The role of the {\it e-e} ({\it h-h}) interaction  has been a
subject of theoretical\cite{Altshuler, Fin, FinRev, Cast1, Cast2,
Cast3, Stern, Gold, Das Sarma} and experimental\cite{Gersh,Bergman,
Wheeler} studies for more than two decades.  The new interest in the
matter is associated with discussion of the nature of metallic-like
temperature dependence of the conductivity observed at low
temperature in some two dimensional (2D) systems, e.g., in n-Si
MOSFET and in dilute 2D hole gas in Al$_x$Ga$_{1-x}$As/GaAs and
Ge$_{1-x}$Si$_x$/Ge structures (see
Refs.~\onlinecite{Pudalov,Savchenko, Shashkin}  and references
therein). As a rule such behavior is observed in low-density
high-mobility structures with the relatively large value of the gas
parameter $r_s =\sqrt{2}/(a_Bk_F)$ characterizing the interaction
strength and by too high value of $T\tau$ (hereafter we set
$\hbar=1$, $k_B=1$), where $a_B$, $k_F$, and $\tau$ are the Bohr
radius, the Fermi quasimomentum, and the transport relaxation time,
respectively. The role of the interaction  at $r_s>3-5$ (i.e., at
strong interaction), and/or  $T\tau\gtrsim 1$ (i.e., at intermediate
and ballistic regimes)  was theoretically studied in
Refs.~\onlinecite{Fin, Fin1, Fin2, Gold, Aleiner1, Aleiner2, Gornyi,
Das Sarma}, the experimental situation was reviewed  in
Refs.~\onlinecite{Pudalov, Savchenko, Shashkin}.

It should be noted that  the metallic-like behavior is observed when
the conductivity  is not too high,  therefore the corrections can
lead to essential change of the conductivity with the temperature.
The changing  of the interaction correction at decreasing
temperature and/or conductivity was theoretically studied in
framework of the theory of the renormalization group (RG) in the
papers.\cite{Cast1,Cast2,Cast3,FinRev,Fin1, Fin2} It has been shown
that the correction renormalization depends on both the Drude
conductivity and the Fermi liquid amplitude $\gamma_2$ that controls
the {\it e-e} interaction in the triplet channel. The contributions
from singlet and triplet channels are opposite in sign favoring
localization and antilocalization, respectively. In conventional
conductors with high values of the Drude conductivity,
$\sigma_0=\pi\, k_Fl\,G_0\gg G_0$ [where $l$ is the mean free path
and $G_0=e^2/(2\pi^2\hbar)$ ], the initial value of the amplitude
$\gamma _2$ is small, and the net effect is in favor of
localization. At $\sigma_0 \lesssim (10-15) G_0 $ or in dilute
systems, however, this amplitude may be enhanced due to {\it e-e}
correlations and thus results in metallic sign of $d\sigma/dT
$.\cite{Fin1, Fin2}

Significantly less is known about the role of the interaction
correction in disordered 2D systems when the $k_Fl$ value tends to
unity, i.e., at crossover from  weak to strong localization.
Experimentally, this effect was studied in the simplest
single-valley electron 2D system GaAs/In$_{1-x}$Ga$_{x}$As/GaAs with
small $\textsl{g}$~factor.\cite{our1} It was shown that  the net
value of the interaction correction decreases rapidly with the
$\sigma_0$ decrease at $\sigma_0 \lesssim (12-15)\,G_0$
($k_Fl\lesssim 4-5$). Such a behavior can  result from the
compensation of the contributions of the singlet and triplet
channels as well as from suppression of  both contributions with
decreasing $\sigma_0$. It is impossible to separate these two
effects in the systems with small value of the  $\textsl{g}$~factor.
The situation changes drastically when dealing with a system with
large enough $\textsl{g}$~factor. In this case the magnetic field
can be used as a tool allowing  to control the ratio between the two
different contributions because it strongly suppresses the triplet
channel and leaves the singlet channel unchanged.  As shown below
the hole 2D gas in strained GaAs/In$_x$Ga$_{1-x}$As/GaAs structures
is a suitable object to study the renormalization of the interaction
quantum correction with the conductivity decrease. In a previous
paper, Ref.~\onlinecite{our2}, we have studied these structures at
high Drude conductivity, $\sigma_0>30\, G_0$.

In this paper we report the results of experimental study of the
evolution of the interaction correction to the conductivity in a
$p$-type 2D system with  decreasing Drude conductivity within the
range from $\simeq 30\,G_0$ to $\simeq 3\,G_0$ when the ballistic
contribution of the  {\it h-h} interaction is small. Firstly, we
will outline the procedures used for extracting the diffusion part
of the interaction correction and the value of the Fermi liquid
parameter $F_0^\sigma=-\gamma_2/\left(1+\gamma_2\right)$ from the
dependences of $\rho_{xx}$ and $\rho_{xy}$ on the temperature and
magnetic field. Then, we will discuss the change of $F_0^\sigma$
with  decreasing Drude conductivity. Finally, we will show that the
reduction of the interaction correction with the decreasing Drude
conductivity results from both the compensation of the singlet and
triplet channels and from the arising  of a prefactor $\alpha_i<1$
in the conventional expression for the interaction
correction.\cite{Fin}

\section{Experiment}

We have measured the temperature and magnetic field dependences of
$\rho_{xx}$ and $\rho_{xy}$ in the heterostructures
GaAs/In$_x$Ga$_{1-x}$As/GaAs grown by metal-organic vapor phase
epitaxy on  semiinsulating  GaAs substrate. The lattice mismatch
between In$_x$Ga$_{1-x}$As and GaAs results in biaxial compression
of the quantum well. The structures consist of a $250$~nm-thick
undoped GaAs buffer layer, carbon $\delta$-layer, a 7~nm spacer of
undoped GaAs, a 10~nm In$_{0.2}$Ga$_{0.8}$As well, a 7~nm spacer of
undoped GaAs, a carbon $\delta$-layer   and 200~nm cap layer of
undoped GaAs. The samples were mesa etched into standard Hall bars
and then an Al gate electrode was deposited by thermal evaporation
onto the cap layer through a mask. Varying the gate voltage $V_g$ we
were able to change the hole density $p$ and mobility $\mu$ within
the following ranges: $p=(2.5\ldots 8.0)\times 10^{11}$ cm$^{-2}$,
$\mu=(1000\ldots 5700)$~cm$^2$/Vs. Two Hall bars prepared from each
of the waffles 3856 and 3857 with close parameters were measured.

\begin{figure}
\includegraphics[width=0.9\linewidth,clip=true]{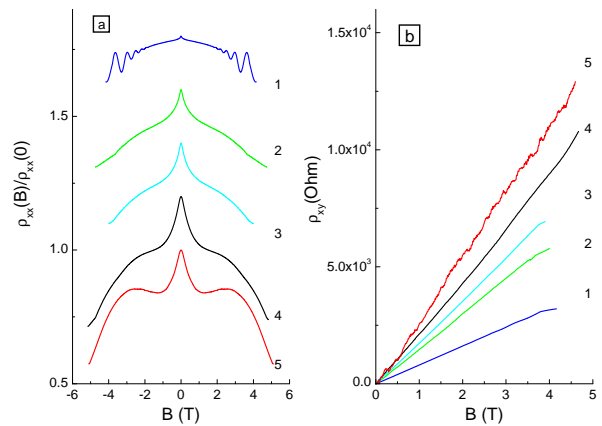}
\caption{The magnetic field dependences of $\rho_{xx}$ (a) and
$\rho_{xy}$ (b) at $T=1.4$ K for different gate voltages, which are
characterized by the following values of $p$, $\sigma_0$, and
$\sigma(T=1.4\,\text{K})$: $8\times 10^{11}$~cm$^{-2}$, $59.6\,G_0$,
and $56.9\,G_0$ (curves 1); $4.5\times 10^{11}$~cm$^{-2}$,
$9.9\,G_0$, and $6.8\,G_0$ (curves 2); $3.9\times
10^{11}$~cm$^{-2}$, $8.1\,G_0$, and $4.37\,G_0$ (curves 3); $3\times
10^{11}$~cm$^{-2}$, $3.9\,G_0$, and $0.36\,G_0$ (curves 4);
$2.6\times 10^{11}$~cm$^{-2}$, $3.5\,G_0$, and $0.027\,G_0$ (curves
5). Structure 3856. For clarity, the curves in the panel (a) are
separated in vertical direction by the value of 0.2. }\label{F1}
\end{figure}

The  magnetic field dependences of $\rho_{xx}$ and $\rho_{xy}$ at
$T=1.4$~K at different gate voltages for one of the samples
investigated are presented in Fig.~\ref{F1}. It is clearly seen that
despite the very large difference in conductivity values at $B=0$,
the magnetoresistance (MR) curves $\rho_{xx}(B)$ are very similar:
the sharp negative MR at low magnetic field, which results from
suppression of the interference contribution to the conductivity, is
followed by the parabolic-like MR caused by the interaction
correction.\cite{Paalanen}

Since our goal is to study the interaction correction let us briefly
explain the method allowing us to extract it from the experimental
data. Under our experimental conditions the parameter $T\tau$ is
small enough ($T\tau<0.1$) and  therefore the main contribution
comes from the diffusion part of the interaction correction. The
unique property of the diffusion part is that it contributes to
$\sigma_{xx}$ but not to $\sigma_{xy}$. This fact opens a
possibility to extract this correction reliably even when the
correction value is small. The most straightforward way is to find
such contribution to $\sigma_{xx}$ which is absent in $\sigma_{xy}$.
We extract these contributions by making use of the structure of the
components of the conductivity tensor $\sigma_{xx}$ and
$\sigma_{xy}$. As shown in Ref.~\onlinecite{our5} the weak
localization correction and the ballistic part of the interaction
corrections are reduced to renormalization of the transport
relaxation time and can be accounted  for  through the temperature
and magnetic field dependence of the mobility. Thus $\sigma_{xx}$
and $\sigma_{xy}$ can be written as
\begin{eqnarray}
\sigma_{xx}(B,T)&=&\frac{ep\mu(B,T)}{1+\mu^2(B,T)B^2}+
\delta\sigma_{xx}^{hh}(B,T), \label{eq3}\\
\sigma_{xy}(B,T)&=&\frac{ep\mu^2(B,T)B}{1+\mu^2(B,T)B^2},
\label{eq4}
\end{eqnarray}
where $\delta\sigma_{xx}^{hh}(B,T)$ is the diffusion part of the
interaction correction. If the Zeeman splitting is very small as
compared with the temperature, $\delta\sigma_{xx}^{hh}$ is magnetic
field independent. It has the form\cite{Fin,Cast1,Cast2,Cast3}
\begin{equation}
\frac{\delta \sigma_{xx}^{hh}(T)}{G_0}=\alpha_i
\left[1+3\left(1-\frac{\ln\left(1+F_0^\sigma\right)}{F_0^\sigma}\right)\right]\ln{T\tau},
\label{eqS0}
\end{equation}
where the first term in square brackets is the exchange or the Fock
contribution while the second one is the Hartree contribution (the
triplet channel). For the following, we enter here the prefactor
$\alpha_i$ which was absent in
Refs.~\onlinecite{Fin,Cast1,Cast2,Cast3}.

\begin{figure}
\includegraphics[width=\linewidth,clip=true]{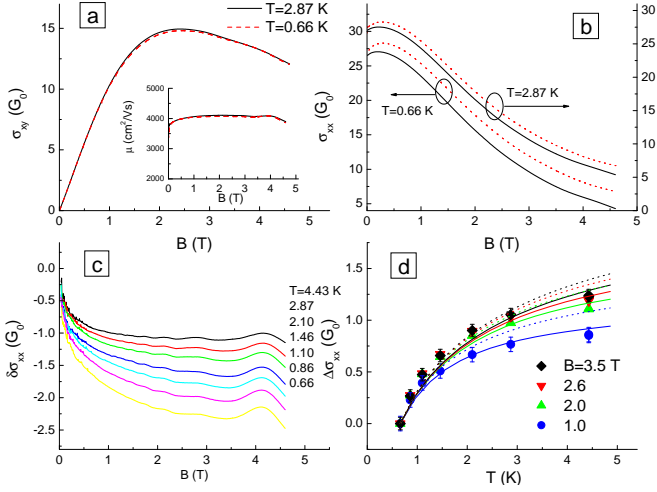}
\caption{(a) The experimental magnetic field dependences of
$\sigma_{xy}$ for two temperatures.  The inset shows the magnetic
field dependence of the mobility calculated from $\sigma_{xy}(B)$
for two temperatures with $p=5.4\times 10^{11}$~cm$^{-2}$. (b) The
$\sigma_{xx}$~vs~$B$ dependences for two temperatures. The solid
curves are the data, the dotted curves are the first term of
Eq.~(\ref{eq3}) calculated as described in text.  (c) The magnetic
field dependences of the difference between experimental and
calculated $\sigma_{xx}$ values for different temperatures. (d) The
temperature dependences of $\Delta\sigma_{xx}$ at different magnetic
fields. The symbols are the experimental results; curves are
calculated dependences with $F_0^\sigma=-0.4$ (solid curves) and
with $F_0^\sigma=-0.35$ (dotted curves). Structure 3857, $V_g=2.4$
V, $\sigma_0\simeq 30\,G_0$.}\label{F2}
\end{figure}

Thus,  knowing the hole density $p$ we can find  $\mu(T,B)$ from
experimental $\sigma_{xy}$~vs~$B$ dependences  [with the help of
Eq.~(\ref{eq4})] and then calculate the first term in
Eq.~(\ref{eq3}). The difference between experimental value of
$\sigma_{xx}$ and this term should give the diffusion part of the
{\it h-h} correction to the conductivity. This method allows us to
find $\delta\sigma_{xx}^{hh}(B,T)$ for relatively low $\sigma_0$,
when the interference contribution to MR is not negligible up to the
high magnetic field.

In what follows we demonstrate how this method works considering the
results obtained for  one of the samples, fabricated on the basis of
structure 3857.

Let us start with the case of  high  Drude conductivity,
$\sigma_0\simeq 30\,G_0$  (for the details of determination of
$\sigma_0$ see Ref.~\onlinecite{our3}).  First for each temperature
we have inverted the resistivity tensor whose components  measured
experimentally and found the conductivity tensor  components
$\sigma_{xy}$ and $\sigma_{xx}$ [solid curves in Figs.~\ref{F2}(a)
and \ref{F2}(b)]. Then, using the obtained $\sigma_{xy}$~vs~$B$
dependences we have found $\mu(B)$  [shown in inset in
Fig.~\ref{F2}(a)] and calculated the experimental value of the first
term in Eq.~(\ref{eq3}). Finally, subtracting  the latter term from
the experimental value of $\sigma_{xx}$ we obtain $\delta
\sigma_{xx}$ [see Fig.~\ref{F2}(c)], which is identified with the
diffusion part of the {\it h-h} correction
$\delta\sigma_{xx}^{hh}(B,T)$. As seen from Fig.~\ref{F2}(b)
$\delta\sigma_{xx}$ is  a small difference between two large
quantities. That is why an accuracy in determination of
$\delta\sigma_{xx}^{hh}(B,T)$, i.e., the absolute value of the
interaction correction, is sufficiently low. In particular, it is
very sensitive to the value of hole density, which is experimentally
known with some accuracy. However, the difference of the quantities
$\delta\sigma_{xx}$ taken at two temperatures for a given magnetic
field (or taken at two magnetic fields for a given temperature)
depends only slightly on the hole density and, therefore, is found
with  better accuracy.

In Fig.~\ref{F2}(d) we present the temperature  dependences $\Delta
\sigma_{xx}(T,B)=\delta \sigma_{xx}(T,B)-\delta \sigma_{xx}(T_0,B)$,
where $T_0$ is the lowest temperature, obtained for different
magnetic field. One can see that the higher is the magnetic field,
the stronger is the change of $\Delta \sigma_{xx}$ with the
temperature. This dependence can be attributed to the Zeeman
splitting which leads to suppression of the triplet channel and,
hence, to  appearance of the magnetic field dependence of the
interaction correction. Theoretically, the effect of Zeeman
splitting has been considered in
Refs.~\onlinecite{Fin3,Cast3,Raimondi90}, and \onlinecite{Cast1}.
However, the expressions derived there are too complicated and,
therefore, inconvenient for the practical use. Much simpler
expression, which well approximates these formulas,
is\cite{Igor,our2}
\begin{eqnarray}
{\delta \sigma^{hh}_{xx}\over G_0}&=&\alpha_i\left\{
\ln{T\tau}+\left[1-
\frac{\ln(1+F_0^\sigma)}{F_0^\sigma}\right]\right.
\nonumber \\
 &\times & \left.\left[\ln{T\tau}+2
\ln T\tau\sqrt{1+\left(\frac{\textsl{g}\mu_B
B}{T}\right)^2}\right]\right\}. \label{eq2}
\end{eqnarray}
In Fig.~\ref{F2}(d) we plot the  curves calculated according to
Eq.~(\ref{eq2}) with $\alpha_i=1$, $\textsl{g}=3$,\cite{ftn1}  and
different $F_0^\sigma$ values. One can see that the curves
calculated with $F_0^\sigma=-0.4$ almost coincide  with the
experimental data.

\begin{figure}
\includegraphics[width=\linewidth,clip=true]{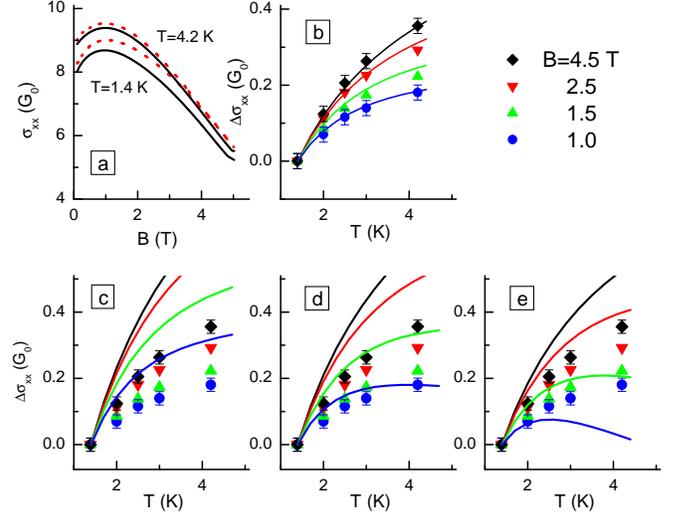}
\caption{(a) The magnetic field dependences of $\sigma_{xx}$ for two
temperatures.  The solid curves are the data, the dotted curves are
the first term of Eq.~(\ref{eq3}) calculated as described in text.
(b) -- (e) The temperature dependences of $\Delta\sigma_{xx}$ at
different magnetic fields. The symbols are the experimental results;
the curves are  theoretical dependences calculated with:
$\alpha_{i}=0.5$, $F_0^\sigma=-0.43$ [panel (b)]; $\alpha_{i}=1$,
$F_0^\sigma=-0.45$ [panel (c)]; $\alpha_{i}=1$, $F_0^\sigma=-0.5$
[panel (d)]; $\alpha_{i}=1$, $F_0^\sigma=-0.55$ [panel (e)].
Structure 3857, $V_g=2.8$~V, $p=4.4\times 10^{11}$~cm$^{-2}$,
$\sigma_0\simeq 11\,G_0$.}\label{F3}
\end{figure}

Similar data treatment was carried out for the lower conductivity.
In Fig.~\ref{F3} we present  the experimental and calculated
magnetic field dependences of $\sigma_{xx}$ [Fig.~\ref{F3}(a)] and
$\Delta\sigma_{xx}$-versus-$T$ dependences for different magnetic
fields [Fig.~\ref{F3}(b)]  for $\sigma_0=11\,G_0$ ($V_g=2.8$~V). As
seen from Figs.~\ref{F3}(c) -- \ref{F3}(e),  it is impossible to
describe the data by Eq.~(\ref{eq2}) with the prefactor $\alpha_i=1$
for any $F_0^\sigma$-values. This is not surprising because the
theory predicts $\alpha_i=1$ only for large $\sigma_0$ value.
However one can fit the data perfectly  with $\alpha_i=0.5$ and
$F_0^\sigma=-0.43$ [see Fig.~\ref{F3}(b)].\cite{fnt2}

\begin{figure}
\includegraphics[width=\linewidth,clip=true]{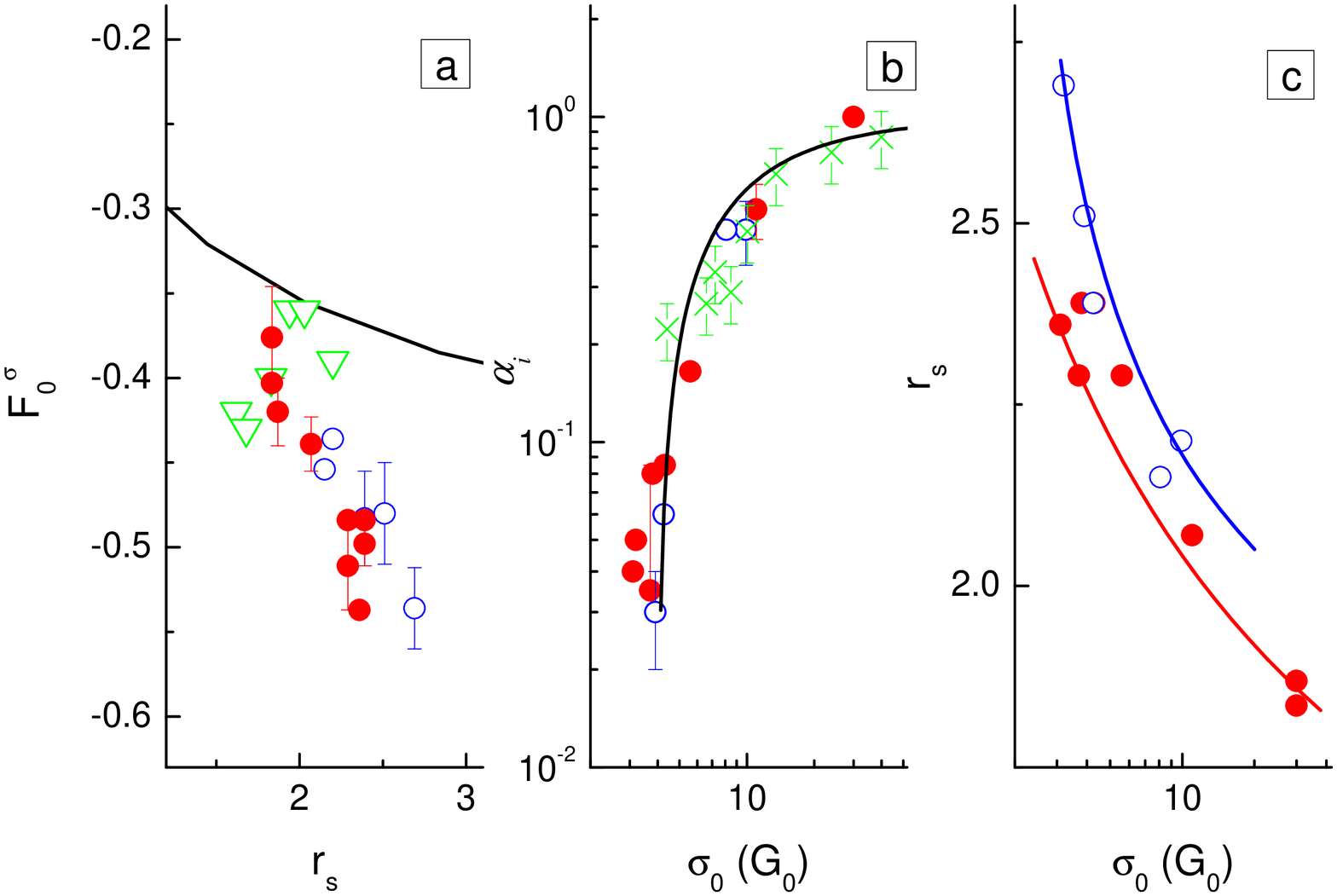}
\caption{(a)  The $r_s$ dependences of $F_0^\sigma$. (b) and (c) The
$\sigma_0$ dependence of the prefactor $\alpha_i$ and the gas
parameter $r_s$, respectively. Open and solid circles are the
experimental data obtained in the present paper for structures 3856
and 3857, respectively. The triangles are data from
Ref.~\onlinecite{our2}. The crosses are the results of recalculation
of the data obtained in Ref.~\onlinecite{our1}. The curve in panel
(a) is theoretical dependence, Eq.~(\ref{eq41}).  The curve in panel
(b) is the interpolating formula, Eq.~(\ref{alphaVSs0}). The curves
in panel (c) are provided as a guide for the eye. }\label{F4}
\end{figure}

To be sure that these changes in $F_0^\sigma$ and $\alpha_i$ are not
random we carried out systematical studies of the both structures at
successive decrease of the hole density and Drude conductivity. It
was recognized that Eq.~(\ref{eq2}) with the two fitting parameters,
$\alpha_i$ and $F_0^\sigma$, describes well the experimental data
down to $\sigma_0\simeq 3.5\pm0.3$.  All the results for $\alpha_i$
and $F_0^\sigma$ are summarized in Fig.~\ref{F4}. The results  of
Ref.~\onlinecite{our2} for $F_0^\sigma$, obtained for $\sigma_0>30\,
G_0$ are presented in Fig.~\ref{F4}(a) also. One can see that all
data match well. The scatter of the data from Ref.~\onlinecite{our2}
is broader than that obtained here due to the large ballistic
contribution that complicates the determination of $F_0^\sigma$.
Note the $\alpha$~vs~$\sigma_0$ data can be interpolated by the
empirical formula
\begin{equation}
 \alpha_i=1-\frac{4\,G_0}{\sigma_0}.
 \label{alphaVSs0}
\end{equation}

Let us firstly discuss the behavior of the Fermi liquid parameter
$F_0^\sigma$. Its value as a function of the gas parameter, $r_s$,
is plotted in Fig.~\ref{F4}(a). It is seen that $F_0^\sigma$
appreciably decreases with   $r_s$  that it becomes less  than
$-0.454$ at $r_s\simeq 2$. It is the value where the interaction
correction in zeroth magnetic field changes sign [see
Eqs.~(\ref{eqS0}) and (\ref{eq2})]. However, the change of the sign
of the interaction correction does not result in the metallic-like
behavior of the total conductivity. This is because the
insulating-like  WL quantum correction dominates in our samples.
Nevertheless, this fact manifests itself in our experiment. Since
the triplet channel is suppressed with the $B$-increase, the
magnetic field inverts the sign of $\delta\sigma_{xx}^{hh}$ again.
So the magnetoresistance should be positive at low magnetic field
and negative at high field. This fact graphically shows itself as
the maximum in $\rho_{xx}$~vs~$B$ dependence, which is evident for
$\sigma_0\simeq 3.5\,G_0$ at $B\simeq 2.8$~T [see Fig.~\ref{F1}(a)].

\begin{figure}
\includegraphics[width=0.7\linewidth,clip=true]{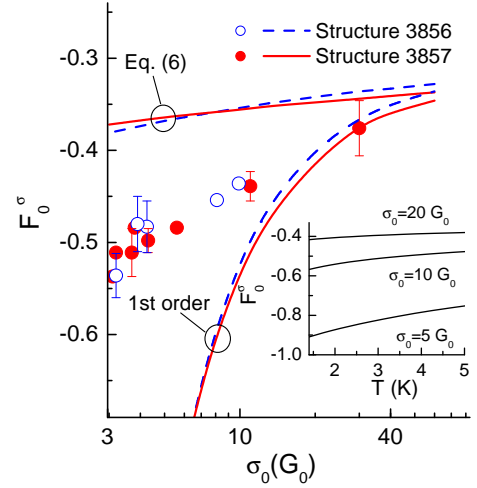}
\caption{(a) The $\sigma_0$ dependence of the interaction parameter
$F_0^\sigma$. The symbols are the experimental data. The curves are
theoretical dependences  calculated as described in text.}\label{F5}
\end{figure}

In  Fig.~\ref{F4}(a) we have plotted the theoretical $r_s$
dependence of $F_0^\sigma$[Ref.~\onlinecite{Aleiner1}]
\begin{eqnarray}
 F_0^\sigma &=& -\frac{1}{2\pi}\frac{r_s}{\sqrt{2-r_s^2}}
  \ln\left(\frac{\sqrt{2}+\sqrt{2-r_s^2}}{\sqrt{2}-\sqrt{2-r_s^2}}\right),\,\,\, r_s^2<2\nonumber \\
 &  &
 -\frac{1}{\pi}\frac{r_s}{\sqrt{r_s^2-2}}\arctan{\sqrt{\frac{1}{2}r_s^2-1}},\,\,\,
 r_s^2>2.
 \label{eq41}
\end{eqnarray}
It is seen that experimental points strongly deviate downwards from
the theoretical curve  with increasing $r_s$. The possible reason of
the deviation is the renormalization of the Fermi liquid constant
$F_0^\sigma$ with the decreasing   Drude conductivity which strongly
changes with $r_s$ [see Fig.~\ref{F4}(c)]. This is directly evident
from Fig.~\ref{F5}, where both the experimental and theoretical
[Eq.~(\ref{eq41})] $F_0^\sigma$ vs $\sigma_0$ dependences are
presented. When calculating the theoretical curves we have used the
$r_s$ vs $\sigma_0$ dependences from Fig.~\ref{F4}(c). It is seen
that the lower the Drude conductivity the stronger the deviation.

Theoretically, the effect of renormalization of $F_0^\sigma$ with
the changing conductivity was studied in the framework of RG
theory,\cite{Cast1,Cast2,Cast3,FinRev,Fin1, Fin2} which took the
interaction into account in the first order in $1/\sigma$ exactly.
According to this theory the temperature dependences of both
$\sigma$ and $F_0^\sigma$ are the solutions of the system of
differential equation
\begin{eqnarray}
\frac{d\sigma}{d\xi}& =&
-\left\{1+1+3\left[1-\frac{1+\gamma_2}{\gamma_2}\ln(1+\gamma_2)\right]\right\}
\label{eq5}\\
\frac{d\gamma_2}{d\xi}&=&\frac{1}{\sigma}\frac{\left(1+\gamma_2\right)^2}{2}
\label{eq6}
\end{eqnarray}
where $\xi=-\ln(T\tau)$,
$\gamma_2=-F_0^\sigma/\left(1+F_0^\sigma\right)$, and $\sigma$ is
measured in units of $G_0$. The term $1+1$ in braces is responsible
for the weak localization and the interaction in singlet channel
which in the case of Coulomb interaction give equal contributions.

The above system of differential equations have been solved
numerically with the following initial conditions. We suppose that
the high-temperature conductivity is equal to the Drude
conductivity: $\sigma(\xi=0)=\sigma_0$. The second condition is
$\gamma_2(\xi=0)=-F_0^\sigma/\left(1+F_0^\sigma\right)$ where
$F_0^\sigma$ is determined by Eq.~(\ref{eq41}).\cite{fnt3} Note this
system describes the conductivity as a function of the parameter
$T\tau$ (i.e., as a function of temperature). Experimentally, we are
able to find the interaction contribution within  only the
relatively narrow temperature range, $T=1.4-4.5$~K. Therefore it is
more appropriate to compare the $\sigma_0$ dependence of
$F_0^\sigma$ rather than the temperature one. The solutions obtained
for several $\sigma_0$ values, as the $F_0^\sigma$~vs~$T$ dependence
within the actual temperature range are presented in inset in
Fig.~\ref{F5}. It is seen that this dependence is relatively weak.
In order to compare these results with the experimental data for
$F_0^\sigma$ we have averaged the calculated value of $F_0^\sigma$
over this temperature interval. Namely this averaged value is
presented in Fig.~\ref{F5} as a function of $\sigma_0$.

It is seen that both the experimental and calculated values
(labelled as ``1st order'') of $F_0^\sigma$ decrease with decreasing
$\sigma_0$. However, the theory gives much faster decrease. Most
probably such a discrepancy indicates that one should take into
account the next terms in $1/\sigma$ expansion in RG equations. To
the best of our knowledge this has been done only for two particular
cases inappropriate to our situation. The first case relates to
multi-valley ($n_v\gg 1$) systems with $\gamma_2\ll 1$.\cite{Fin2}
The second one is single valley ($n_v= 1$) systems but with the
large $\gamma_2$ value.\cite{Kirkpatr90}

Thus, realizing the crudity of the above estimations we,
nevertheless, believe that the decrease of the experimental value of
$F_0^\sigma$ with the decreasing Drude conductivity results from the
renormalization of the {\it h-h} interaction.

Strictly speaking, the RG equations (\ref{eq5}) and (\ref{eq6}) were
derived in the absence of magnetic field, whereas the Zeeman
splitting suppresses the triplet contributions to the right-hand
side of Eqs.~(\ref{eq5}) and
(\ref{eq6}).\cite{Fin3,Raimondi90,Cast1,Cast3} Recently, it was
shown that the effect of Zeeman splitting on conductivity can be
used for extracting the dependence of $F_0^\sigma$ on
temperature.\cite{Anis06,Burmistr06,Knyaz06} In our case the
temperature range in which the Zeeman splitting is strong,
$\textsl{g}\mu_B B>T$, is small fraction of the total interval which
we use for averaging. Therefore, we do not expect significant
difference in our results for $F_0^\sigma$ due to taking into
account the Zeeman splitting and consider the comparison of our data
with solutions of Eqs.~(\ref{eq5}) and (\ref{eq6}) is almost
correct.

Next we discuss  the behavior of the prefactor $\alpha_i$.
Fig.~\ref{F4}(b) shows that $\alpha_i$ decreases sharply when
$\sigma_0$ lowers. The behavior of the interaction correction with
decreasing $\sigma_0$ was studied experimentally for the $n$-type 2D
structures in Ref.~\onlinecite{our1}. The recalculated data from
this paper presented in Fig.~\ref{F4}(b) by crosses demonstrate
analogous decrease also. The possible reason of such
$\alpha_i$~vs~$\sigma_0$ dependence is the interplay  between the
interference and the interaction which has not been taken into
account in the RG theory.\cite{Fin1,Fin2} As shown in
Refs.~\onlinecite{Aleiner3} and \onlinecite{our4} two additional
terms in the expression for the conductivity arise if this interplay
is allowed for [see Eq.~(40) in Ref.~\onlinecite{our4}]. One term
depends on the magnetic field and leads to appearance of the
prefactor in WL magnetoresistance. The second one does not depend on
the magnetic field, and therefore it was away in
Ref.~\onlinecite{our4}. It is quite possible that namely this term
leads to decrease of $\alpha_i$, prefactor in the interaction
correction, with decreasing Drude conductivity. Another contribution
to the prefactor $\alpha_i$ is due to the second-loop interaction
effect. This correction is known for the singlet channel in the
unitary ensemble (strong magnetic field).\cite{Burmistr02} To the
best of our knowledge the impact of the interplay between the
interaction and the interference upon the interaction correction to
the conductivity as well as the second-loop contribution in the
triplet channel for $n_v=1$ is yet to be studied.

In summary, the behavior of the interaction contribution to the
conductivity with decreasing Drude conductivity is determined both
by the renormalization of the interaction constant $F_0^\sigma$ and
by the decrease of the prefactor $\alpha_i$ in Eq.~(\ref{eq2}), and
the latter is more pronounced.

\subsection*{Acknowledgments}
We would like to thank I.~V. Gornyi for very useful discussions and
I.~S. Burmistrov for a critical reading of the manuscript and
valuable comments. This work was supported in part by the RFBR
(Grant Nos. 05-02-16413, 06-02-16292, and 07-02-00528), the CRDF
(Grant No. Y3-P-05-16).


\begin{thebibliography}{}
\bibitem{Altshuler} B.~L. Altshuler and A.~G. Aronov, in {\it Electron-Electron
Interaction in Disordered Systems}, edited by A.~L.~Efros and
M.~Pollak, (North Holland, Amsterdam, 1985) p.1.

\bibitem{Fin} A.~M. Finkelstein, Zh. Eksp. Teor. Fiz. \textbf{84}, 168 (1983) [Sov.
Phys. JETP \textbf{57}, 97 (1983)]; Z. Phys. B \textbf{56}, 189
(1984).

\bibitem{FinRev} A. M. Finkel'stein, {\it Electron Liquid in Disordered Conductors},
Vol.~14 of {\it Soviet Scientific Reviews}, edited by I. M.
Khalatnikov (Harwood, London, 1990).


\bibitem{Gold} A.~Gold and V.~T. Dolgopolov, Phys. Rev. B \textbf{33}, 1076 (1986).

\bibitem{Das Sarma} S.~Das Sarma and E.~H. Hwang, Phys. Rev. Lett. \textbf{83}, 164 (1999).


\bibitem{Cast1} C.~Castellani, C.~Di Castro, P.~A. Lee, and M.~Ma, Phys. Rev. B
\textbf{30}, 527 (1984).

\bibitem{Cast2} C.~Castellani, C.~Di Castro, P.~A. Lee, and M.~Ma, Phys. Rev. B
\textbf{30}, 1596 (1984).

\bibitem{Cast3}  C.~Castellani, C.~Di Castro, P.~A. Lee, Phys. Rev. B \textbf{57}, R9381
(1998).


\bibitem{Stern} F.~Stern, Phys. Rev. Lett. \textbf{44}, 1469 (1980).


\bibitem{Gersh} B.~L. Altshuler, A.~G. Aronov, M.~E. Gershenson, and Yu.~V. Sharvin,
Sov. Sci. A. Phys. \textbf{9}, 223 (1987).

\bibitem{Bergman} G.~Bergman, Physics Reports, \textbf{107}, 1 (1984).

\bibitem{Wheeler} K.~M. Cham and R.~G. Wheeler, Phys. Rev. Lett. \textbf{44}, 1472 (1980).

\bibitem{Pudalov} V.~M. Pudalov, M.~E. Gershenson, and H. Kojima,
in {\it Fundamental Problems of Mesoscopic Physics: Interactions and
Decoherence}, edited by I.~V. Lerner, B.~L. Altshuler, and Y. Gefen
(Kluwer Academic Publishers, Dordrecht, 2004), p.~309; E.~A.
Galaktionov, A.~K. Savchenko, S.~S. Safonov, Y.~Y. Proskuryakov, L.
Li, M. Pepper, M.~Y. Simmons, D.~A. Ritchie, E.~H. Linfield, and
Z.~D. Kvon, {\it ibid}. p.~349.

\bibitem{Savchenko} A.~K. Savchenko, Y.~Y. Proskuryakov, S.~S. Safonov, L.~Li, M.~Pepper, M.~Y. Simmons, D.~A.
Ritchie, E.~H. Linfield, and Z.~D. Kvon, Phisica E \textbf{22}, 218
(2004).

\bibitem{Shashkin}  A.~A. Shashkin, Uspekhi Fiz. Nauk \textbf{175}, 139 (2005) [Physics-Uspekhi \textbf{48}, 129
(2005)].

\bibitem{Fin1} Alexander Punnoose, Alexander M. Finkel'stein,
Phys. Rev. Lett. \textbf{88}, 016802 (2001).

\bibitem{Fin2} Alexander Punnoose, Alexander M. Finkel'stein, Science \textbf{310}, 289
(2005).


\bibitem{Aleiner1} Gabor Zala, B.~N. Narozhny, and I.~L. Aleiner, Phys. Rev. B {\bf 64},
214204 (2001).

\bibitem{Aleiner2} Gabor Zala, B.~N. Narozhny, and I.~L. Aleiner, Phys. Rev. B {\bf 64}, 201201 (2001).

\bibitem{Gornyi} I.~V. Gornyi  and A.~D. Mirlin, Phys. Rev. Lett. \textbf{90}, 076801
(2003); Phys. Rev. B \textbf{69}, 045313 (2004).

\bibitem{our1} G.~M. Minkov, O.~E. Rut, A.~V. Germanenko, A.~A.
Sherstobitov, V.~I. Shashkin, O.~I. Khrykin, and B.~N. Zvonkov,
Phys. Rev. B {\bf 63}, 205306 (2003).

\bibitem{our2} G.~M. Minkov, O.~E. Rut, A.~V. Germanenko, A.~A.
Sherstobitov, V.~A. Larionova, and B.~N. Zvonkov, Phys. Rev. B {\bf
72}, 165325 (2005).

\bibitem{Paalanen}M.~A. Paalanen, D.~C. Tsui, and J.~C.~M. Hwang, Phys. Rev. Lett.
{\bf 51}, 2226 (1983).

\bibitem{our5} G.~M. Minkov, A.~V. Germanenko, O.~E. Rut, A.~A. Sherstobitov,
V.~A. Larionova, A.~K. Bakarov, and B.~N. Zvonkov, Phys. Rev. B {\bf
74}, 045314 (2006).

\bibitem{our3} G.~M. Minkov, A.~V. Germanenko, O.~E. Rut, A.~A.
Sherstobitov, and B.~N. Zvonkov, arXiv:~cond-mat/0606566
(unpublished).


\bibitem{Fin3} A. M. Finkel'stein, Zh. Eksp. Teor. Fiz. {\bf 86},
367 (1984) [Sov. Phys. JETP {\bf 59}, 212 (1984)].

\bibitem{Raimondi90} R. Raimondi, C. Castellani, and C.~Di Castro, Phys. Rev. B {\bf 42}, 4724
(1990).

\bibitem{Igor} I.~V. Gornyi, private communication.


\bibitem{ftn1} In our previous paper\cite{our2} we have used $\textsl{g}=3$. The parameters
of the interaction correction depend weakly on specific value of
$\textsl{g}$~factor, however $\textsl{g}=3$ gives the best agreement
over whole conductivity range.


\bibitem{fnt2} Note the hole density is relatively high under our experimental
condictions, $p\gtrsim 2.5\times 10^{11}$~cm$^{-2}$, and we believe
that the renormalization of $\textsl{g}$~factor due to the hole-hole
interaction is negligible.


\bibitem{fnt3} It should be mentioned
that the final result presented in Fig.~\ref{F5} is qualitatively
independent of the concrete initial value of $F_0^\sigma$ when it
lies within the reasonable interval $F_0^\sigma=-(0.55\ldots 0.45)$.
In all the cases the calculated curve strongly deviates down from
the experimental data  when $\sigma_0 \lesssim 10\, G_0$.


\bibitem{Kirkpatr90} T.~R. Kirkpatrick and D. Belitz, Phys. Rev. B \textbf{41}, 1108
(1990).

\bibitem{Anis06} S.~Anissimova, S.~V. Kravchenko, A.~Punnoose, A.~M. Finkelstein, T.~M. Klapwijk,
arXiv:~cond-mat/0609181 (unpublished).

\bibitem{Burmistr06}  I.~S. Burmistrov and N.~M. Chtchelkatchev, JETP
Lett. \textbf{84}, 656 (2006).

\bibitem{Knyaz06} D.~A. Knyazev, O.~E. Omel'yanovskii, V.~M.
Pudalov, and I.~S. Burmistrov, JETP Lett. \textbf{84}, 662 (2006).

\bibitem{Aleiner3}I.~L. Aleiner, B.~L. Altshuler, and M.~E. Gershenzon, Waves Random Media \textbf{9}, 201 (1999).

\bibitem{our4} G.~M. Minkov, A.~V. Germanenko and I.~V. Gornyi, Phys. Rev.
B {\bf 70}, 245423 (2004).

\bibitem{Burmistr02} M.~A. Baranov, I.~S. Burmistrov, and A.~M.~M. Pruisken, Phys. Rev. B \textbf{66},
075317 (2002).

\end{thebibliography}
\end{document}